\documentclass[nofootinbib,aps,superscriptaddress,reprint,floatfix]{revtex4}
\pdfoutput=1
\usepackage{etoolbox}
\makeatletter
\patchcmd{\maketitle}{\@fpheader}{}{}{}
\makeatother

\usepackage{graphicx,epsf,bm}
\usepackage{amssymb}
\usepackage{amsmath}
\usepackage{epstopdf}
\usepackage{booktabs}
\usepackage{subfigure}
\usepackage{rotating}
\usepackage{xspace}

\newcommand{\heptt}{HEPTopTagger\xspace}
\newcommand{\http}{HEPTopTagger$'$\xspace}
\newcommand{\hptt}{HPTTopTagger\xspace}
\newcommand{\pythia}{PYTHIA\xspace}
\newcommand{\herwigpp}{HERWIG++\xspace}
\newcommand{\htt}{\heptt}
{}
{}
{}
\newcommand{\W}{\ensuremath{W}\xspace}
\newcommand{\pt}{\ensuremath{p_{T}}\xspace}
\newcommand{\ttbar}{\ensuremath{t\bar{t}}\xspace}
\newcommand{\Zprime}{\ensuremath{{Z'}}\xspace}
\newcommand{\CamKt}{\ensuremath{\mathrm{C/A}}\xspace}
\newcommand{\delphes}{Delphes\xspace}
\newcommand{\fastjet}{FastJet\xspace}
\newcommand{\figref}[1]{Figure~\ref{fig:#1}}
\newcommand{\secref}[1]{Section~\ref{sec:#1}}
\renewcommand{\eqref}[1]{Equation~(\ref{eq:#1})}
\newcommand{\tabref}[1]{Table~\ref{tab:#1}}

\begin{document}

\title{\bf \Large Tagging highly boosted top quarks}

\author{\large S. Sch\"atzel}
\affiliation{Physikalisches Institut, Ruprecht-Karls-Universit\"at Heidelberg, Germany}

\author{\large  M. Spannowsky}
\affiliation{Institute for Particle Physics Phenomenology,
  University of Durham, Durham DH1 3LE, UK.}

\begin{abstract}
For highly energetic top quarks, the products of the decay $t\rightarrow b q q'$
are collimated. The 3-prong decay structure can no longer be resolved using calorimeter
information alone if the particle jet separation approaches the calorimeter granularity.
We propose a new method, the \hptt,\footnote{Source code available from \url{http://www.ippp.dur.ac.uk/~mspannow/webippp/HPTTopTagger.html}} that
uses tracks of charged particles inside a fat jet to find top quarks with
transverse momentum $\pt > 1$~TeV.
The tracking information is complemented by the calorimeter measurement of the fat
jet energy to eliminate the sensitivity to jet-to-jet fluctuations in the charged-to-neutral particle ratio.
We show that with the \hptt, a leptophobic narrow-mass \Zprime boson of mass 3~TeV
could be found using 300~fb$^{-1}$ of 14~TeV LHC data.
\end{abstract}

\def\thepage{{}}
\maketitle
\def\thepage{\arabic{page}}

\section{Introduction}
\label{sec:case}

After the recent discovery of the Higgs
boson~\cite{Aad:2012tfa,Chatrchyan:2012ufa} the LHC's next foremost goal is to find
evidence for physics beyond the Standard Model, i.e., new particles or forces. With a
center-of-mass energy of 13~TeV, starting in spring 2015, the LHC experiments
can access an unprecedented energy regime, allowing for the production of very heavy resonances.
When heavy TeV-scale resonances decay to electroweak-scale particles
(e.g., top quarks, $W$, $Z$, and Higgs bosons), these particles
are boosted, and for central production, the particle's transverse momentum $\pt$ exceeds its
mass $m$. The decay products of these particles are then collimated in the laboratory frame.
Due to the large branching ratio of electroweak-scale resonances into jets, it
is beneficial, in many measurements and searches, to use jet
substructure methods to disentangle the signal from large QCD
backgrounds~\cite{mike,butterworth,bdrs,boost2010,subreview,boost2011}.
The reconstruction of intermediately ($2 m \leq p_{\rm T}^{} < 5 m$)
and highly ($p_{\rm T}^{} \geq 5 m$) boosted top quarks was one of the
first motivations to study jet substructure
techniques~\cite{had_resonances,semi_resonances}. For an overview of these
so-called `top-taggers' see~\cite{topreview}.

\begin{figure}[ht]
\includegraphics[width=3.0in]{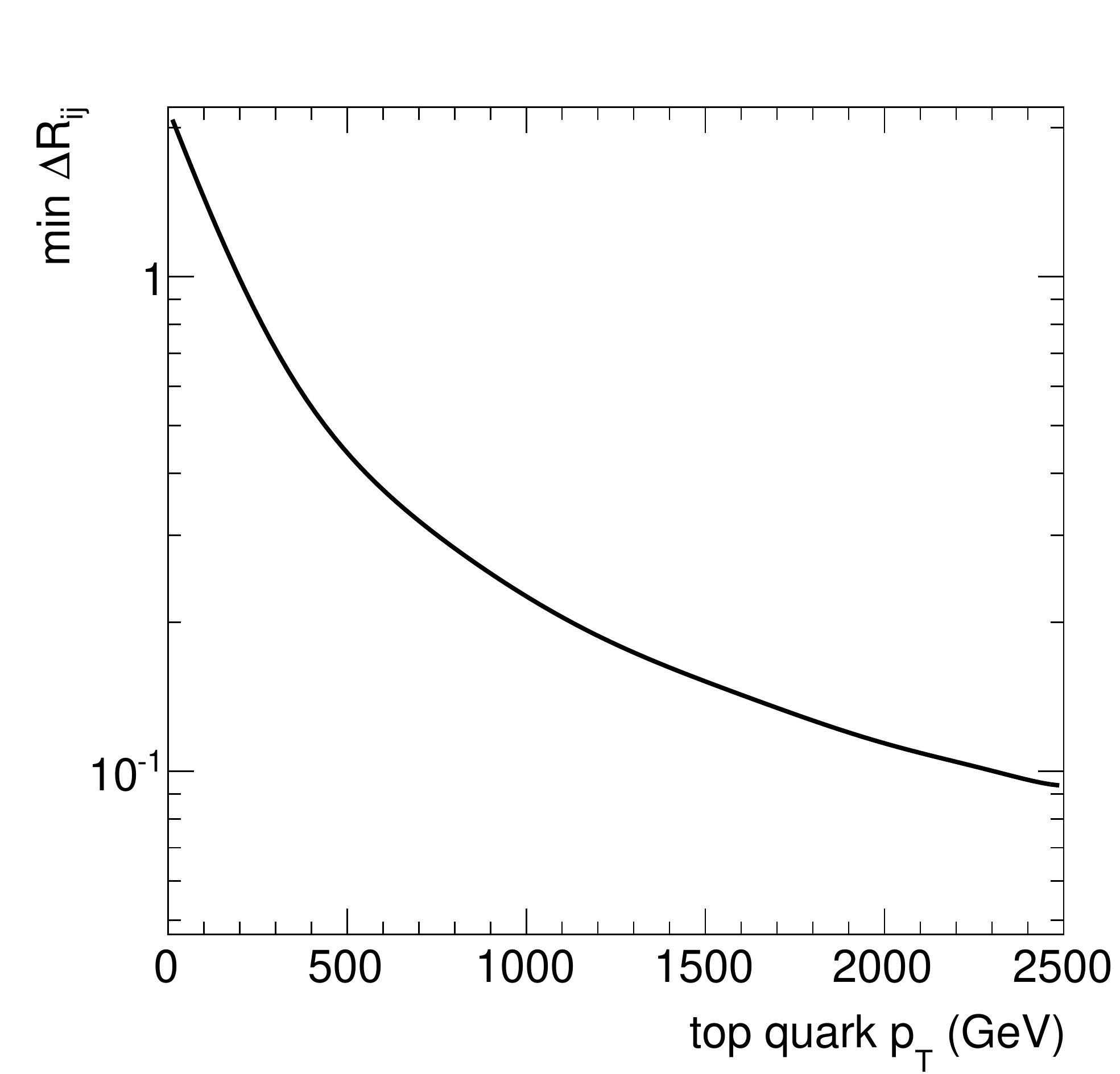}
\caption{Angular separation $\Delta R = \sqrt{(\Delta\eta)^2+(\Delta\phi)^2}$
of the two closest quarks in the top quark decay $t\rightarrow b q q'$ as a
function of the top quark \pt.}
\label{fig:minSeparation}
\end{figure}

Some of the most successful taggers are either based on jet-shape observables~\cite{thaler_wang, leandro1, nsub,treeless} or on sophisticated ways of
selecting subjets inside a fat jet and comparing their energy
sharing~\cite{hopkins,pruning1, cmstagger,heptop, Soper:2012pb}. The separate identification of the decay
products of highly boosted top quarks becomes experimentally challenging
when the detector granularity does not allow to resolve the individual
particle jets.\footnote{For similar problems in decays of electroweak gauge bosons see~\cite{Katz:2010mr}.} This is particularly an issue if jets are reconstructed using
calorimeter information alone as it is currently done with ATLAS data. The
cell size of the ATLAS barrel hadronic calorimeter is $0.1\times 0.1$ in
$(\eta,\phi)$ and topological cell clusters are formed around
seed cells with an energy $|E_{\rm cell}|>4\sigma_{\rm noise}$ by adding
the neighboring cells with $|E_{\rm cell}|>2\sigma_{\rm noise}$, and then
all surrounding cells~\cite{Aad:2012vm}.
The minimal transverse size for a cluster of hadronic calorimeter cells is therefore
$0.3\times0.3$ and is reached if all significant activity is
concentrated in one cell. Two particle jets leave
distinguishable clusters if each jet hits only a single cell
and the jet axes are separated by at least $\Delta R=0.2$,
so that there is one empty cell between the two seed cells.\footnote{A splitting algorithm
has to be used in this case to divide this big cluster into two.}
If two top quark decay jets are so close that they
do not leave separate clusters then top taggers based on identifying
the 3-prong decay structure will fail. \figref{minSeparation} shows the angular
separation $\Delta R = \sqrt{(\Delta\eta)^2+(\Delta\phi)^2}$ of the two
closest final state quarks in hadronic top quark decay $t\rightarrow b q q'$ as a function of the
top quark \pt. For $\pt = 1.12$~TeV the separation is $0.2$ and
calorimeter resolution issues should become apparent around that \pt or even
earlier if the particle level jets that correspond to the quarks are
not collimated enough to hit only a single cell.
A precise determination of the threshold top quark \pt requires the use of the
ATLAS simulation framework and hence has to be carried out by the
ATLAS collaboration.

Tracking detectors that measure the trajectories of charged particles can
remedy the problem because of the much finer spatial resolution.
In this article, we therefore propose a novel method of reconstructing highly boosted
top quarks using a combination of tracks and calorimeter information. We
compare the tagging efficiency of this high-\pt top tagger (\hptt) with the
\htt~\cite{heptop} and show that, with the \hptt, the LHC has a discovery reach
for heavy resonances, which decay exclusively into top quarks, up to a resonance mass of
3~TeV with $300$~fb$^{-1}$ of data taken at $\sqrt{s}=14$~TeV.

The article is arranged as follows:
In \secref{toptag} we introduce the top-tagging algorithm of the \hptt.
In \secref{perf} we compare the tagging performance of the \hptt with
that of the \htt for highly boosted top quarks. We present the reach of the
LHC in discovering very heavy resonances in \secref{res} and summarize our findings
in \secref{outlook}.

\section{Highly boosted top quark reconstruction}
\label{sec:toptag}

The \heptt uses a mass-drop criterion and adaptive filtering to obtain
three subjets that are tested for kinematic compatibility with hadronic top quark decay.
These conditions are formulated in the form of ratios of invariant mass combinations
of the subjets. For example, the mass $m_{23}$ is defined as the invariant mass
of the subleading \pt and the sub-subleading \pt subjet.
For most hadronic top quark decays, the ratio $m_{23}/m_{123}$ corresponds to $m_W/m_t$
with the $b$-jet having the largest \pt. The invariant masses are determined from
the 4-momenta of the subjets which have to be reconstructed precisely.
The ATLAS collaboration has calculated calibration constants for the
\heptt subjets and uncertainties that quantify to which precision the subjet
energy scale and energy resolution can be modeled in simulation~\cite{Aad:2013gja}.
These uncertainties are crucial for comparing the data to simulated model predictions and for setting exclusion
limits as has been done for example in~\cite{Aad:2012raa}.
Calibrations of ATLAS
Cambridge/Aachen (\CamKt) jets~\cite{ca_algo}
are available for radius parameters $R$ from
1.2 down to 0.2~\cite{Aad:2013gja}.
Jets with a smaller radius parameter approach the minimal hadronic cluster size
as discussed in \secref{case}.

A tracking detector can reconstruct the trajectory of a charged particle
and can specify the direction of the particle at any point of the trajectory
with a precision much better than the granularity of the calorimeter.
For example, the angular resolution of the ATLAS inner tracking detector for
charged particles with $\pt = 10$~GeV and $\eta = 0.25$
is $\approx 10^{-3}$ in $\eta$ and $\approx 0.3$~mrad in $\phi$~\cite{Aad:2008zzm} with a reconstruction efficiency of
$>78\%$ for tracks of charged particles with $\pt>500\,{\rm MeV}$~\cite{Aad:2010ac}.
The momentum resolution for charged pions is $4\%$ for momenta $p<10$~GeV,
rising to $18\%$ at $p=100$~GeV~\cite{Aad:2008zzm}.

Prong-based tagging algorithms usually require the reconstruction of the top
quark and \W boson masses to identify a fat jet as being initiated by a
top quark decay. In a typical proton-proton collision, about $65\%$ of the
jet energy is carried by charged hadrons, $25\%$ by photons, produced mainly
from $\pi^0$ decays, and only $10\%$ by neutral hadrons (mostly neutrons and
$K^0_L$)~\cite{energyrat}. However, these
fractions can vary significantly from event to event.
Thus, reconstructing the correct resonance
mass is a challenging task for a tagging algorithm which is based
exclusively on tracks. Fortunately, while no calibrations exist for subjets with
$R<0.2$, the energy of the fat jet can be calibrated to
good precision~\cite{Aad:2013gja} and the inverse of the
energy fraction carried by charged tracks

\begin{equation}
\alpha_j = \frac{E_{\mathrm{jet}}}{E_{\mathrm{tracks}}}
\label{eq:alpha}
\end{equation}
can be measured for each jet individually, thereby eliminating the sensitivity to fluctuations to a large extend.

Our tagger for highly boosted top quarks uses elements of the \htt
which do not introduce artificial mass scales in background events, i.e., we
do not consider all possible three subjet combinations until we find a
top-like structure. Such drastic measures might be necessary when the small
boost of the top quark requires to use a very large jet cone to capture all decay
products. In the case of a highly boosted top quark, the decay products are
confined to a small area of the detector and the amount of additional
radiation inside a \CamKt jet with $R=0.8$ is
usually not excessive.\footnote{The amount of additional radiation in a fat jet strongly
depends on the cone size \cite{Dasgupta:2007wa} but also on the overall hadronic activity of
the event and on the color flow of the underlying hard interaction~\cite{Ask:2011zs}.}

To reconstruct highly boosted top quarks we propose the following procedure, labeled for later reference as HPTTopTagger algorithm:
\begin{enumerate}
\item define a jet $j$ using the C/A algorithm with $R=0.8$ from calorimeter clusters.
\item \label{step2} take the tracks with $\pt > 500$~MeV that are associated with $j$ and recombine them to a track-based jet $j_c$.
\item calculate $\alpha_j$ of \eqref{alpha} using $j$ and $j_c$.
\item apply the mass drop procedure introduced in \cite{heptop}: undo the last clustering of the track-based jet $j_c$ into two  subjets $j_{c1},j_{c2}$ with $m_{j_{c1}} > m_{j_{c2}}$. We require $m_{j_{c1}} <
  0.8~m_{j_c}$ to keep $j_{c1}$ and $j_{c2}$. If this condition does not hold we keep only $j_{c1}$. Each subjet $j_{ci}$ we further decompose unless $m_{j_{ci}} < 20~\mathrm{GeV}$. The remaining subjets we add to the list of relevant substructures.
\item if we find fewer than two remaining subjets we consider the tag to have failed. Else, we take the constituents of all subjets surviving the mass drop procedure and multiply their momenta by $\alpha_j$ each.

\item \label{step6} we take all the rescaled constituents and filter them
with resolution $R_{\mathrm{filt}}=\max (0.05,\min(\Delta R_{ij}/2))$, in which $i$ and $j$
run over all remaining subjets after the mass drop procedure. We recombine
all constituents of the 4 hardest filtered subjets and require the resulting
invariant mass to be in a mass window around the top quark mass. We call
this object our top candidate.

\item \label{step7} again we follow the HEPTopTagger and construct exactly
three $p_T$ ordered subjets $j_1, j_2, j_3$ from the top candidate's
constituents. If the masses $(m_{12},m_{13},m_{23})$ satisfy the so-called
A-cut of~\cite{heptop}, we consider the top tag to be successful.

\end{enumerate}
\medskip

One guiding principle of the outlined algorithm is not to bias the mass
distribution for the top and \W candidates. Particularly at high $p_T$,
splittings of massless quarks and gluons can geometrically induce a large
jet mass  $m_j^2 \sim p_{T,j}^2 \Delta R^2_{j_1,j_2}$ \cite{bib:jetmass}.
Depending on the jet \pt cut in the event, the average jet mass can be much
bigger than the top quark mass. Therefore, looking explicitly for structures
in very hard jets which fulfill simple mass requirements can result in a
large fake rate.

\begin{figure}[ht]
\includegraphics[width=3.0in]{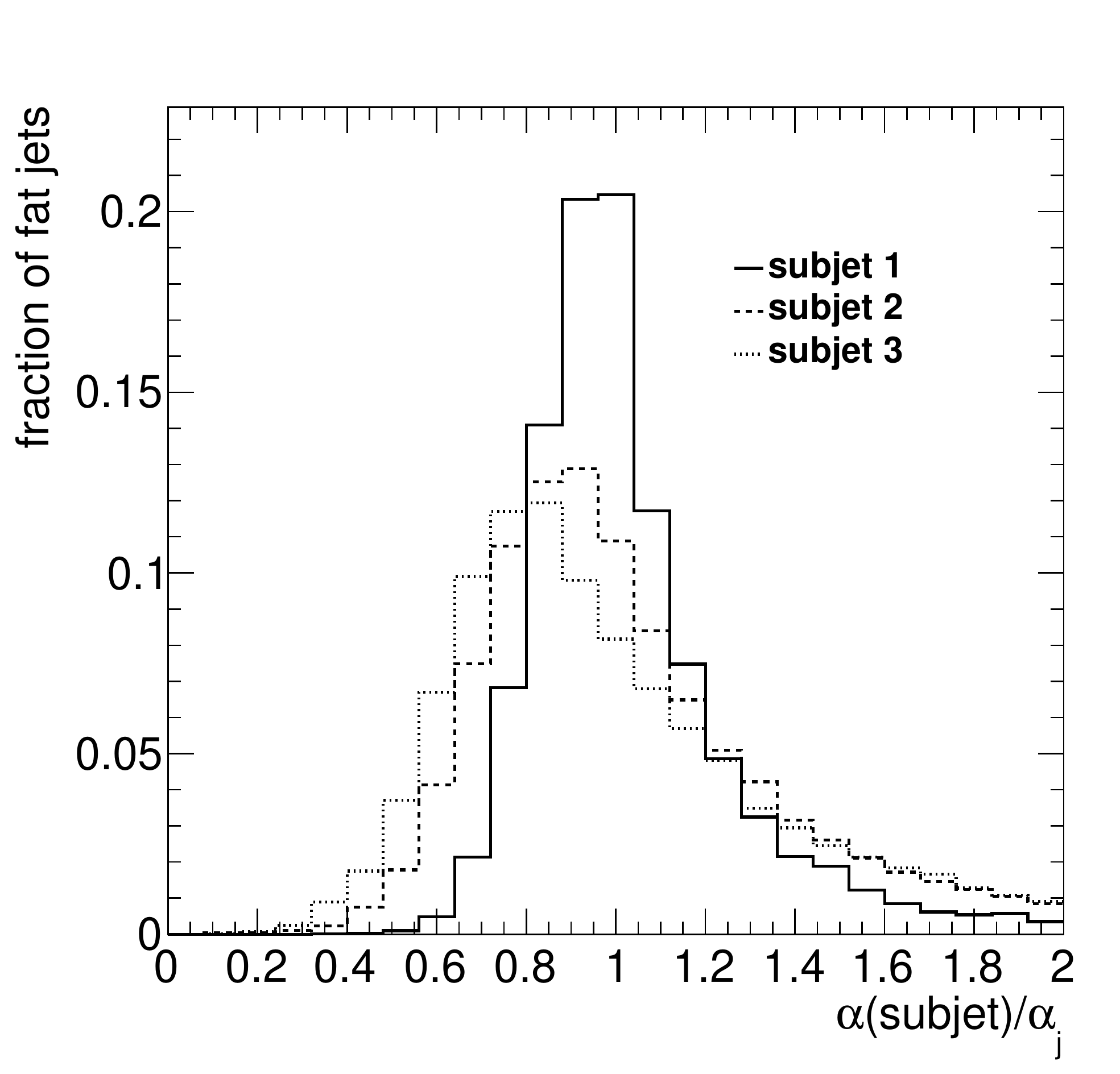}
\caption{The ratio $\alpha({\rm subjet})/\alpha_j$ for the three subjets found
by the \hptt in fat jets in events with $Z'\rightarrow t\bar{t}$ decays where $m_{\Zprime} = 3$~TeV.
Subjet 1 is the leading \pt subjet and subjet 2 the subleading \pt subjet.
}
\label{fig:alpha_subjet}
\end{figure}

The sensitivity to fluctuations in the fraction of charged particles
is reduced by scaling the subjet momenta by $\alpha_j$. This procedure
relies on the assumption that the energy fraction carried by charged particles is
similar in all subjets. \figref{alpha_subjet} compares the particle level subjet
$\alpha$, i.e., the inverse of the energy fraction carried by charged particles with $\pt>500$~MeV
and $|\eta|<2.5$ inside a subjet, to the particle level $\alpha_j$ calculated from the fat jet.
The events used for this figure contain decays of the \Zprime boson of \secref{res}
to $t\bar{t}$ with $m_{\Zprime} = 3$~TeV.
The distributions are very similar for $m_{\Zprime} = 5$~TeV.
For most of the leading \pt subjets (subjets 1), the ratio is $\approx\!0.95$,
whereas for most of the subjets 2 (subleading \pt subjets) and subjets 3, the ratios are $\approx\!0.9$
and $\approx\!0.8$, respectively.
The fat jet quantity $\alpha_j$ is therefore a good approximation to the subjet $\alpha$.
We note that the subjet $\alpha$ cannot be calculated at the detector level
because no calibrations exist for the small calorimeter subjets ($R<0.2$)
we are interested in.

\begin{figure}[ht]
\subfigure[]{
   \includegraphics[width=3.0in]{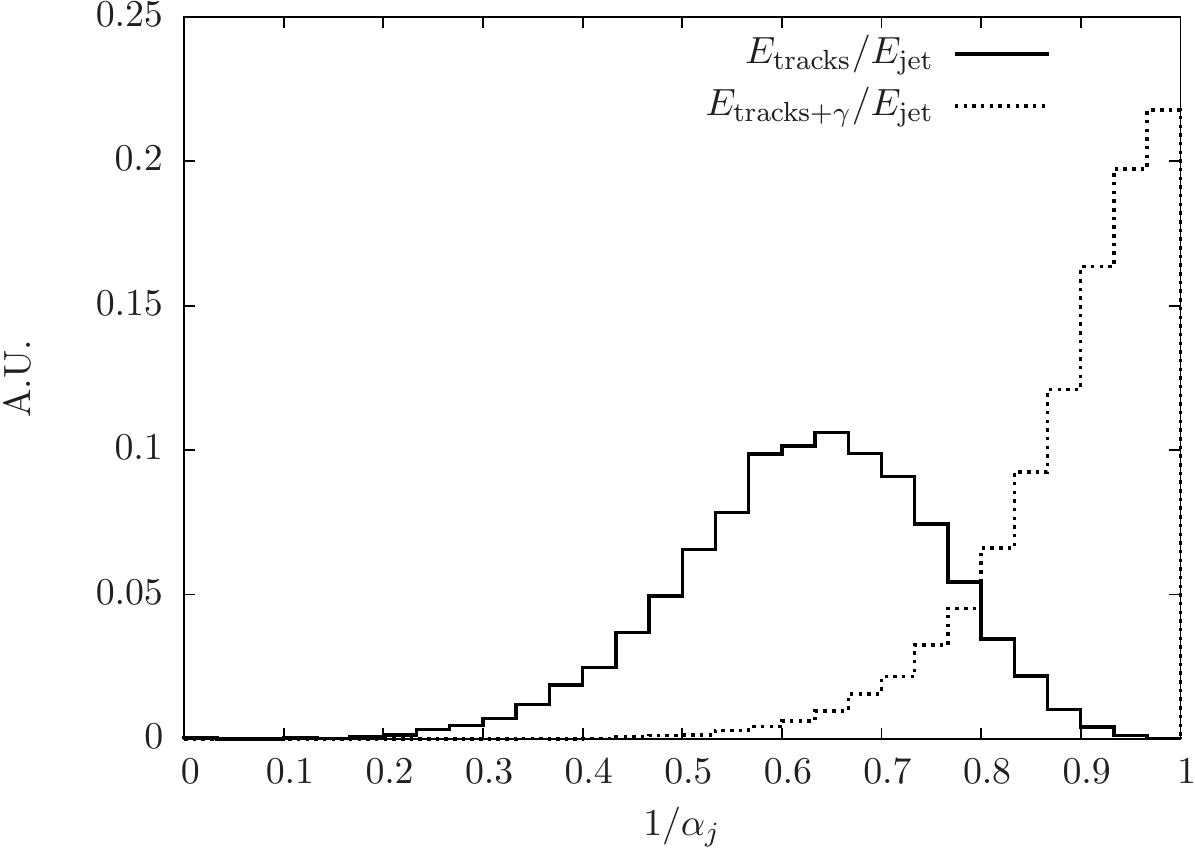}
}
\subfigure[]{
   \includegraphics[width=3.0in]{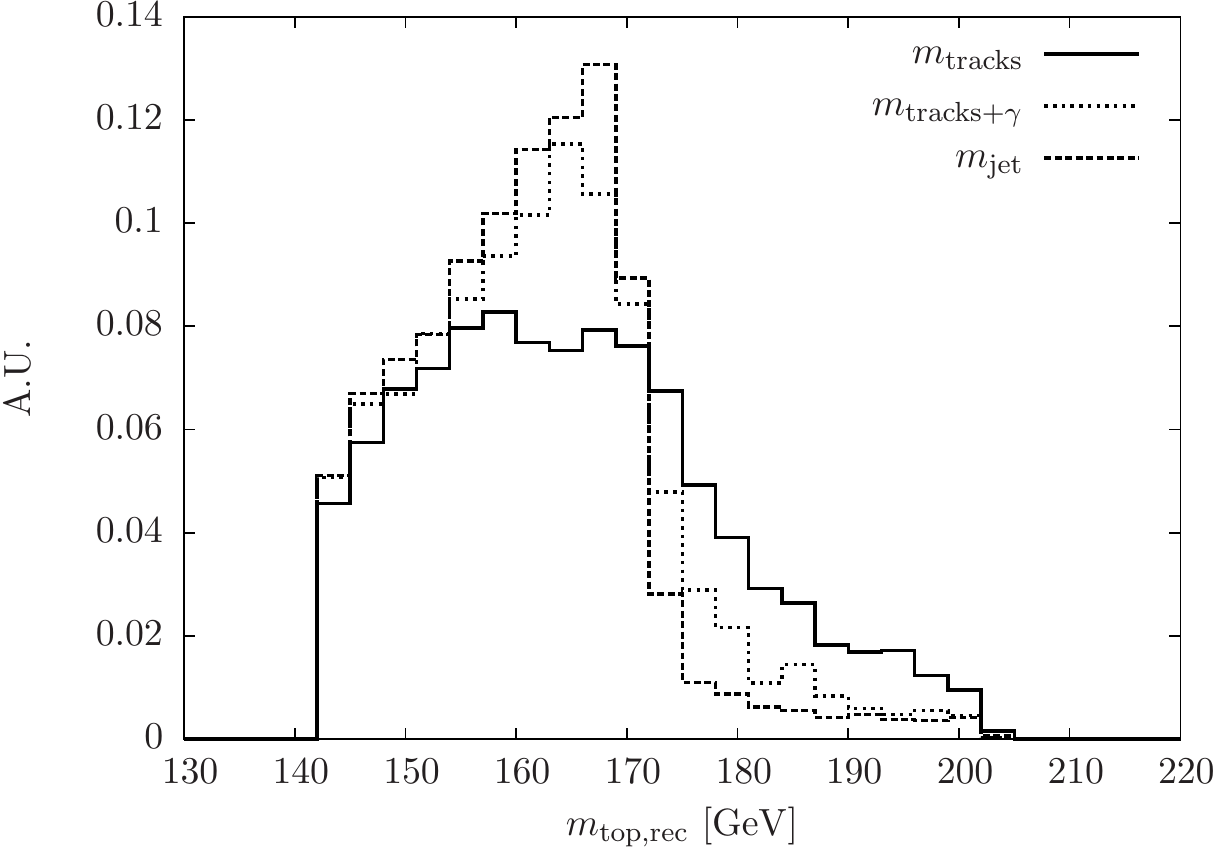}
}
\caption{a) The fat jet energy fraction carried by charged particles with $\pt>500$~MeV ($E_{\rm tracks}/E_{\rm jet}$)
and by charged particles with $\pt>500$~MeV plus photons.
b) The top quark candidate mass reconstructed at the particle level using
the \hptt as defined in \secref{toptag} ($m_{\rm tracks}$) and when
adding photons ($m_{\rm tracks+\gamma}$) or all particles ($m_{\rm jet}$) in step \ref{step2}.}
\label{fig:photons}
\end{figure}

In step \ref{step2} we use only tracks for $j_c$. By
including photons measured in a finely grained electromagnetic calorimeter
in addition to the tracks, it is possible to obtain a richer jet
substructure and a better energy resolution of the rescaled track jet, i.e.,
a better mass reconstruction of the top candidate. This enhancement is currently
not implemented because the fast detector simulation we are using applies the
same segmentation to the electromagnetic and hadronic parts of the calorimeter.
In \figref{photons} we illustrate the impact of adding photon information.
Shown in panel (a) is the fat jet energy fraction carried by charged particles (with $\pt>500$~MeV) and photons
and by the charged particles alone. The distribution is wider in the latter case
because of fluctuations in the photon fraction. The effect on the reconstructed
top mass is shown in panel (b). Here the distribution obtained at the particle level
when applying the above prescription and using only charged particles in step \ref{step2}
is compared to the two cases in which charged particles and photons or all particles are used.
The impact of the fluctuations in the charged-to-neutral particle ratio are already
significantly reduced when adding photon information, leading to a narrower mass distribution.
This better top quark momentum reconstruction will also improve the \ttbar resonance mass resolution.
The sensitivity to the charged-to-neutral fluctuations can be reduced by choosing a large enough
top quark mass window. With a window from $140$ to $210$~GeV we see only a small efficiency increase of a couple
of percent when adding photons.

It is well known that track-based observables are not infrared safe~\cite{chargedIR}. Therefore non-perturbative contributions have to be taken into account to obtain finite and well-defined results. In full event generators like \pythia, hadronization models including fragmentation functions are used. These functions are non-perturbative objects following perturbative evolution equations and are usually fitted to LEP data. The fraction of charged particles is unknown within limits imposed by measurements of the jet fragmentation function~\cite{Aad:2011sc}.  In Section~\ref{sec:perf} we compare the top quark reconstruction efficiency for the HPTTopTagger between \herwigpp and \pythia 8.

Although the cone size for highly boosted top quarks does not need to be
big, we find that using filtering \cite{bdrs, Dasgupta:2013ihk} in step~\ref{step6} improves the
performance of the tagger in separating top jets from QCD jets.
Our goal is to achieve a flat tagging efficiency independent of the top
quark's transverse momentum. Thus, to decide if a tag was successful we use
invariant masses in step~\ref{step7}.

\section{Performance of top-tagging algorithms}
\label{sec:perf}

We use \pythia 8.175~\cite{Sjostrand:2007gs} to obtain fully showered and
hadronized final states of Standard Model \ttbar and dijet production as well
as events with hypothetical leptophobic \Zprime bosons. The \delphes
program~\cite{Ovyn:2009tx} is used in version 2.0.3 to obtain a fast simulation
of the response of an LHC detector. We use the \delphes ATLAS detector card with
a tracking efficiency of $78\%$.
Ultimately, at very high top quark \pt, the track reconstruction will struggle to resolve the
tracks left by nearby particles. The limit is reached when the hits are so close that they are part
of the same reconstructed track and the track reconstruction efficiency suffers as a consequence.
The ATLAS tracking efficiency is $80\%$ for $\pt = 500$~MeV and rises to $86\%$ for $\pt>10$~GeV~\cite{Aad:2010ac,Aad:2010rd}.
To take the close-by effect into account we have used a reduced tracking efficiency of $78\%$ which corresponds
to a $10\%$ relative loss of efficiency at high \pt. We assume that this efficiency is a conservative
lower limit and treat it as constant in \pt. A careful study of the tracking efficiency as a function of \pt is needed but can only be
done with access to the full detector simulation. This is beyond the scope of this article.

The smallest simulated calorimeter entities are calorimeter cells, which
for $|\eta|\leq 2.5$ have a size $0.1\times 0.17$ in $(\eta,\phi)$ (and double this $\phi$ size for $|\eta|>2.5$).
No clustering of these cells is performed, leading to smaller calorimeter structures than
those that would be available with the real ATLAS calorimeter.
The resolution power of the calorimeter is therefore overestimated and the
impact of tracking information will be larger in reality.
For jet finding we use the \fastjet~\cite{fastjet} program.

\CamKt $R=0.8$ jets with $\pt>800$~GeV are built from calorimeter cells and
are required to lie within $|\eta|<2.5$. Tracks with
$\pt>500$~MeV and $|\eta|<2.5$ are matched to these jets using ghost
association~\cite{Cacciari:2007fd, Cacciari:2008gn} as follows. A ghost of
every track is created by setting the \pt to a small value ($10$~eV) and using
the track $\eta$ and $\phi$ at the calorimeter surface. The energy of the ghost
is set to $1.001$ times its momentum to ensure a positive ghost mass. The ghost tracks are added to the
calorimeter jet clustering but don't change the jet because their energy
is negligible. If the ghost track ends up in the jet
then the original track is taken to be associated with the jet. We then cluster
all associated tracks into a \CamKt jet. The calorimeter jet and the track jet
serve as inputs $j$ and $j_c$ to the HPTTopTagger procedure defined in
\secref{toptag}.

The HEPTopTagger as proposed in~\cite{heptop} was designed to work for
mildly boosted top quarks, which required a large radius parameter
of $R=1.5$ to geometrically catch the decay products.
For the reconstruction of highly boosted top quarks we use as inputs to the
\htt the calorimeter $R=0.8$ jets to compare directly to the \hptt but note
that the \htt is optimized to achieve a
high rejection of background that is picked up by using the large radius.
We also modify the original algorithm by stopping the mass drop procedure
already if the subjet mass is below 50~GeV (originally 30~GeV) because this
was the preferred value in~\cite{Aad:2012raa}.

ATLAS subjet calibrations and subjet simulation uncertainties exist for radius
parameters down to $R=0.2$~\cite{Aad:2013gja}.
To demonstrate the performance of the \htt if only those jets were to be used,
we implement the following changes to the original algorithm and
refer to the modified tagger as \http in the following:
\begin{itemize}
\item The minimal filter radius is set to 0.2.
\item Each exclusively clustered subjet is required to have $R'>0.2$ in which
$R'$ is given by the distance in $(\eta, \phi)$ space between the jet axis
and the jet constituent the farthest away from the axis.
\item The filtered subjets and the exclusive subjets must have $\pt>20$~GeV.
\item We stop the mass drop procedure if the two parent jets are closer than $\Delta R = 0.2$.
\end{itemize}

\begin{figure}[ht]
\subfigure[]{
   \includegraphics[width=3.0in]{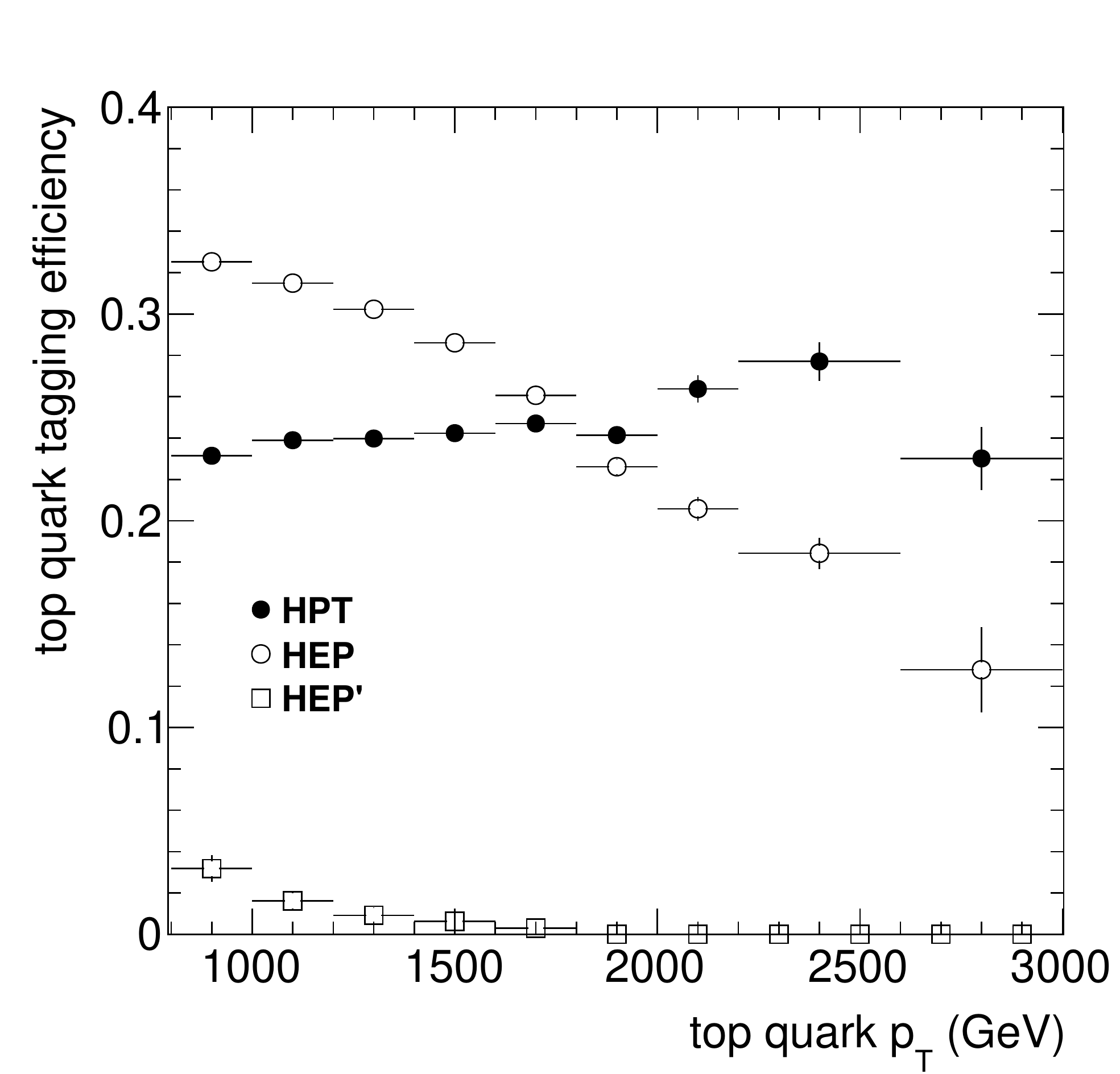}
}
\subfigure[]{
   \includegraphics[width=3.0in]{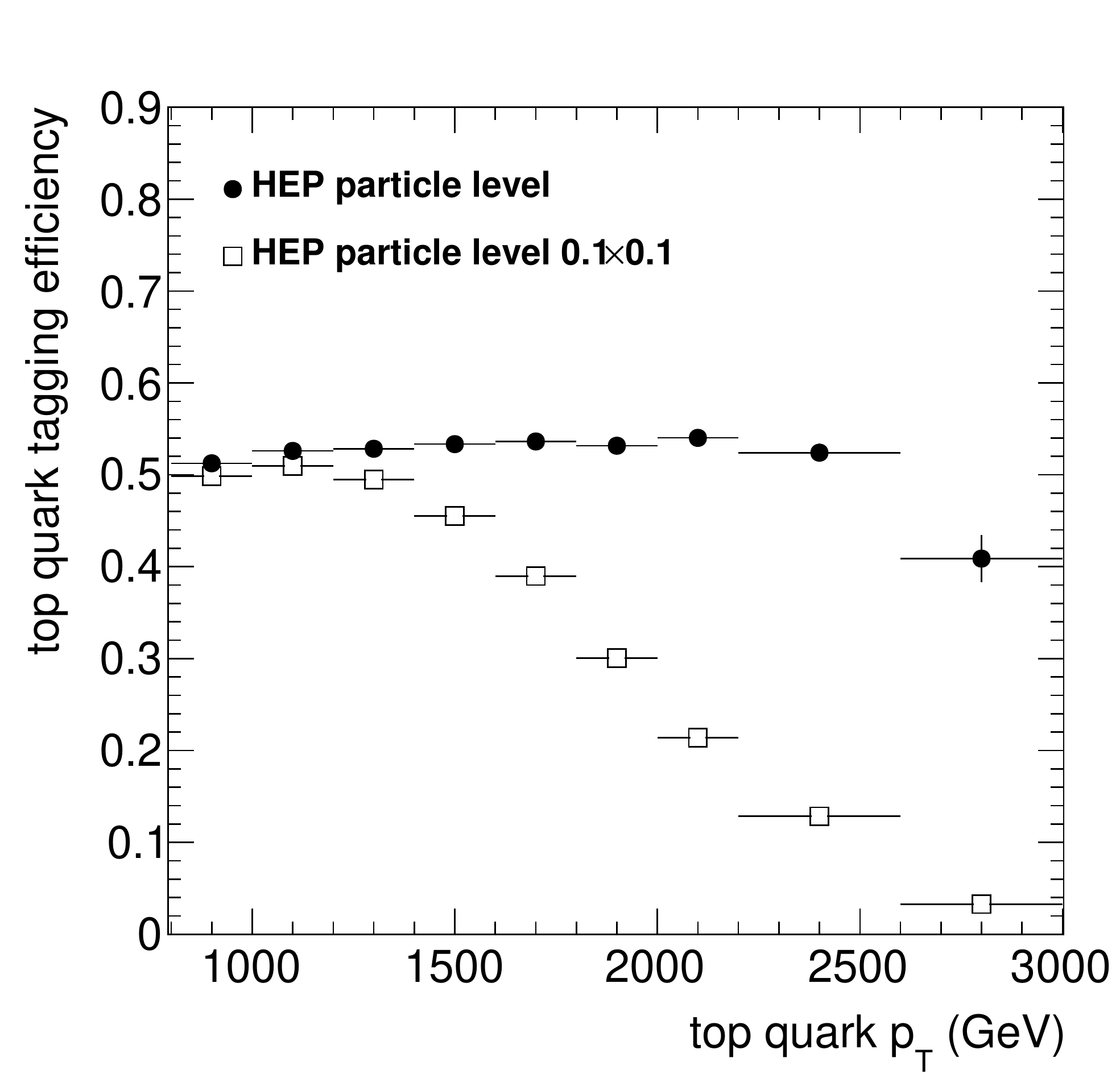}
}
\caption{Efficiencies for tagging top quarks using a) calorimeter cells and b)
stable particles.}
\label{fig:efftagging}
\end{figure}

The same calorimeter fat jets are fed to the three different taggers for \ttbar (signal)
and light quark or gluon dijet events (background).
The top quark tagging efficiency is shown in \figref{efftagging}a as a function of the
top quark \pt. A top quark is taken to be tagged if a reconstructed top quark
candidate is found within $\Delta R=0.6$ of the top quark.
The \hptt efficiency is stable at $\approx 24\%$ up to 3~TeV in \pt.
If the tracking efficiency is artificially set to $100\%$, then the efficiency rises to $28\%$.
This rather small sensitivity to the tracking inefficiency is explained by the fact
that the energy in lost tracks is recovered by the $\alpha_j$ scaling.

For the \htt, the efficiency drops from $\approx 32\%$ for $800<\pt<1000$~GeV to $\approx 13\%$
for $2600<\pt<3000$~GeV due to the segmentation of the calorimeter which prevents
all three top quark decay particle jets from being reconstructed.
To prove this claim, we show in \figref{efftagging}b the top quark finding
efficiency for the \htt at the particle level. When the constituents of the
\CamKt $R=0.8$ jets are stable particles, the \htt efficiency is stable
at $53\%$. If we granularize the particles into $(\eta, \phi)$ cells of size
$0.1\times 0.1$, the efficiency starts to drop at a top quark \pt of $1.2$~TeV.

\begin{figure}[ht]
\includegraphics[width=3.0in]{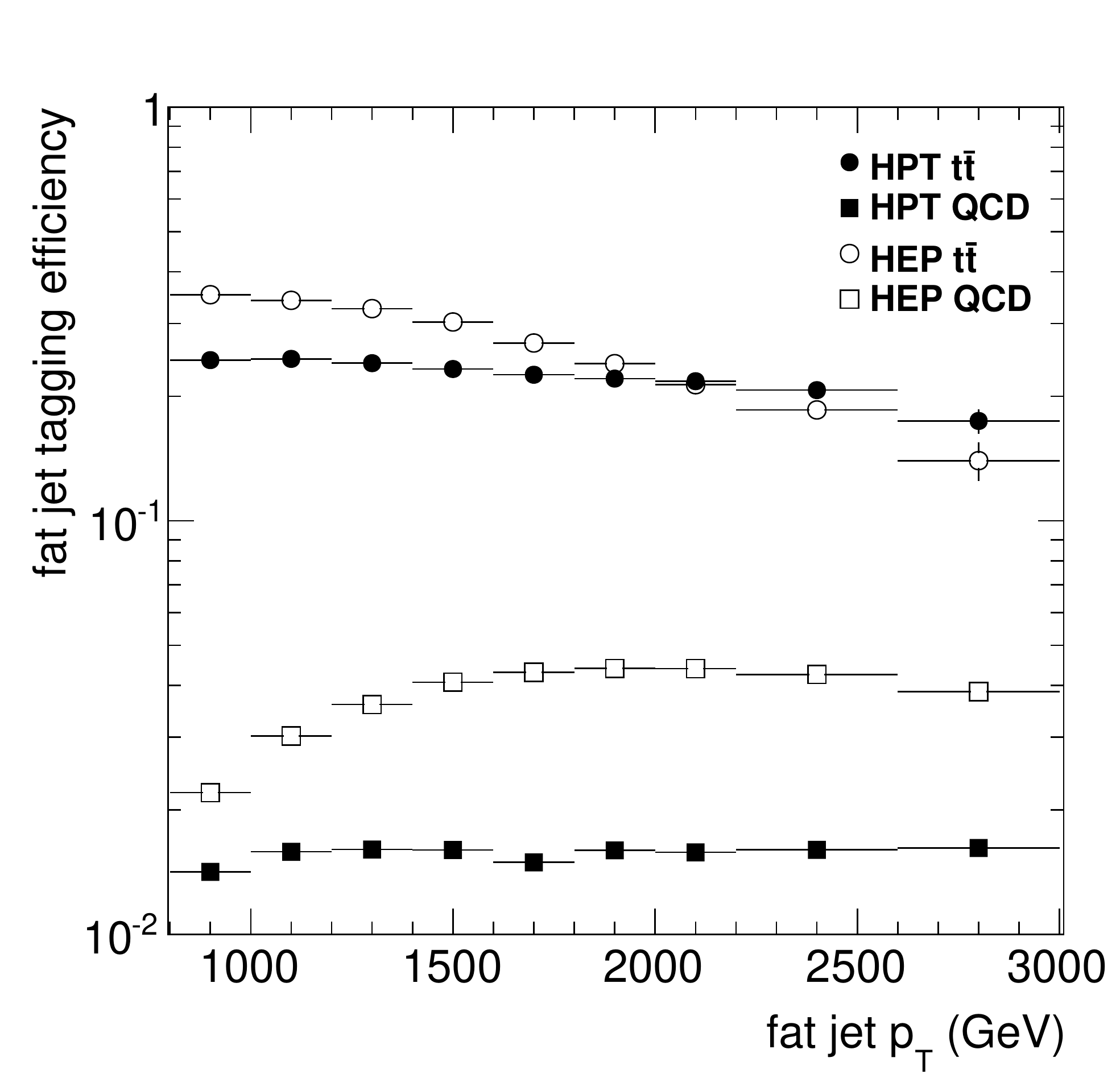}
\caption{Efficiencies for tagging \CamKt $R=0.8$ calorimeter fat jets.}
\label{fig:eff_fj}
\end{figure}

The efficiency of the \http for finding top quarks using calorimeter
cells is less than $4\%$ for $\pt>800$~GeV (\figref{efftagging}a). From this
we conclude that with the present available ATLAS jet calibrations and uncertainties
it is not possible to find top quarks at high \pt. To obtain calibrations and uncertainties
also for jets with $R<0.2$ we suggest the use of the reconstructed top mass peak
in \ttbar{} events. The position of the peak can be used for calibration and the
difference between simulation and data can serve to estimate the simulation uncertainty.
We note that at higher top quark \pt the fraction of subjets with small $R$
will be higher. This effect can be studied by binning the mass distribution
in \pt of the top candidate.

The efficiency for tagging fat jets constructed from calorimeter cells is shown in
\figref{eff_fj} as a function of the fat jet \pt. For \ttbar{} events, the numerical values
are similar to the top quark tagging efficiencies.
The fake rate, defined as the probability to tag fat jets originating from
light quarks or gluons, is stable at $1.6\%$ for the \hptt while it increases
for the \htt from $\approx 2\%$ for $\pt=800$~GeV to $4.5\%$ for $\pt=2$~TeV.
Because of the comparable signal efficiency and the much lower fake rate, the
\hptt outperforms the \htt when the resonance search strategy requires an
improvement on the signal-to-background ratio.
The fake rate is smaller with the \hptt because not all possible three subjet combinations
are tried when looking for a top-like structure. This is different in the \htt where
all triplets of substructure objects (the objects after the mass drop) are
tested for compatibility with top quark decay, which increases the efficiency
but also the fake rate.

\begin{figure}[ht]
\subfigure[]{
   \includegraphics[width=3.0in]{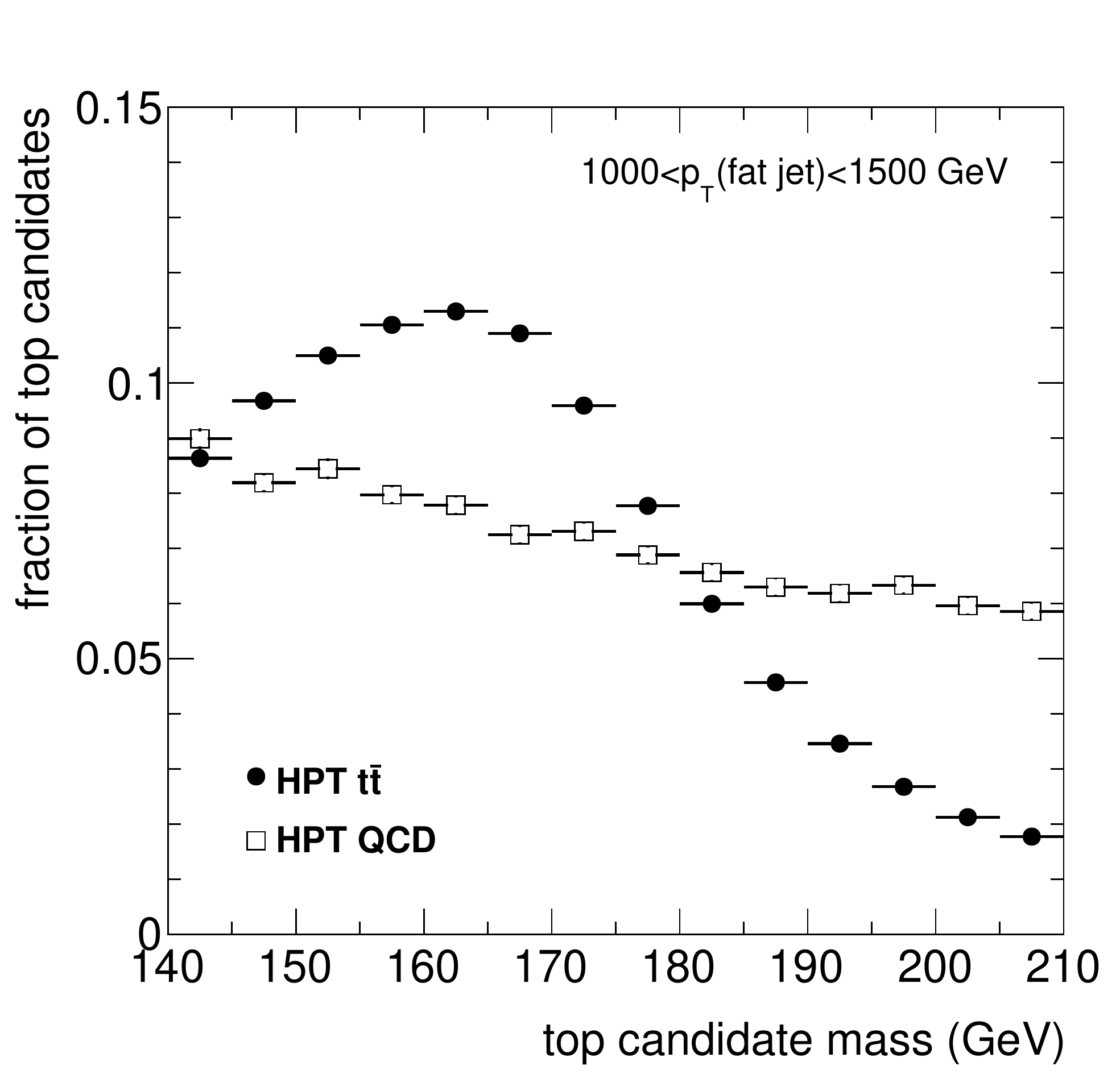}
}
\subfigure[]{
   \includegraphics[width=3.0in]{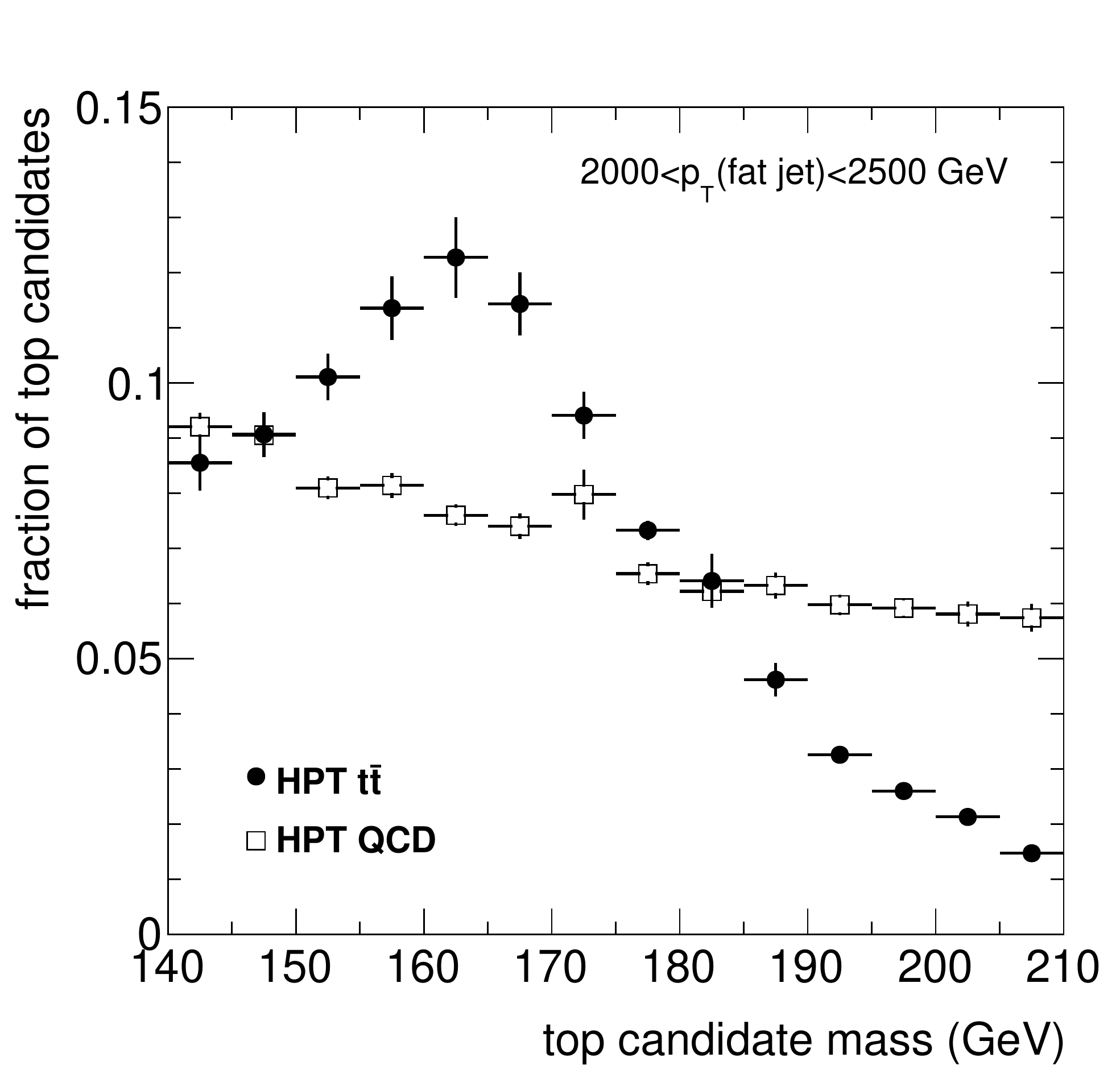}
}
\caption{Top quark mass reconstructed with the \hptt in two bins of the calorimeter fat jet \pt.}
\label{fig:topmass}
\end{figure}

The top quark mass reconstructed with the \hptt is shown in \figref{topmass} in two bins of
the calorimeter fat jet \pt. While a clear peak is visible for events with top quarks,
the background distribution is smoothly falling and shows no shaping into a peak.
This holds true for low and high fat jet \pt.

\begin{figure}[ht]
\includegraphics[width=3.0in]{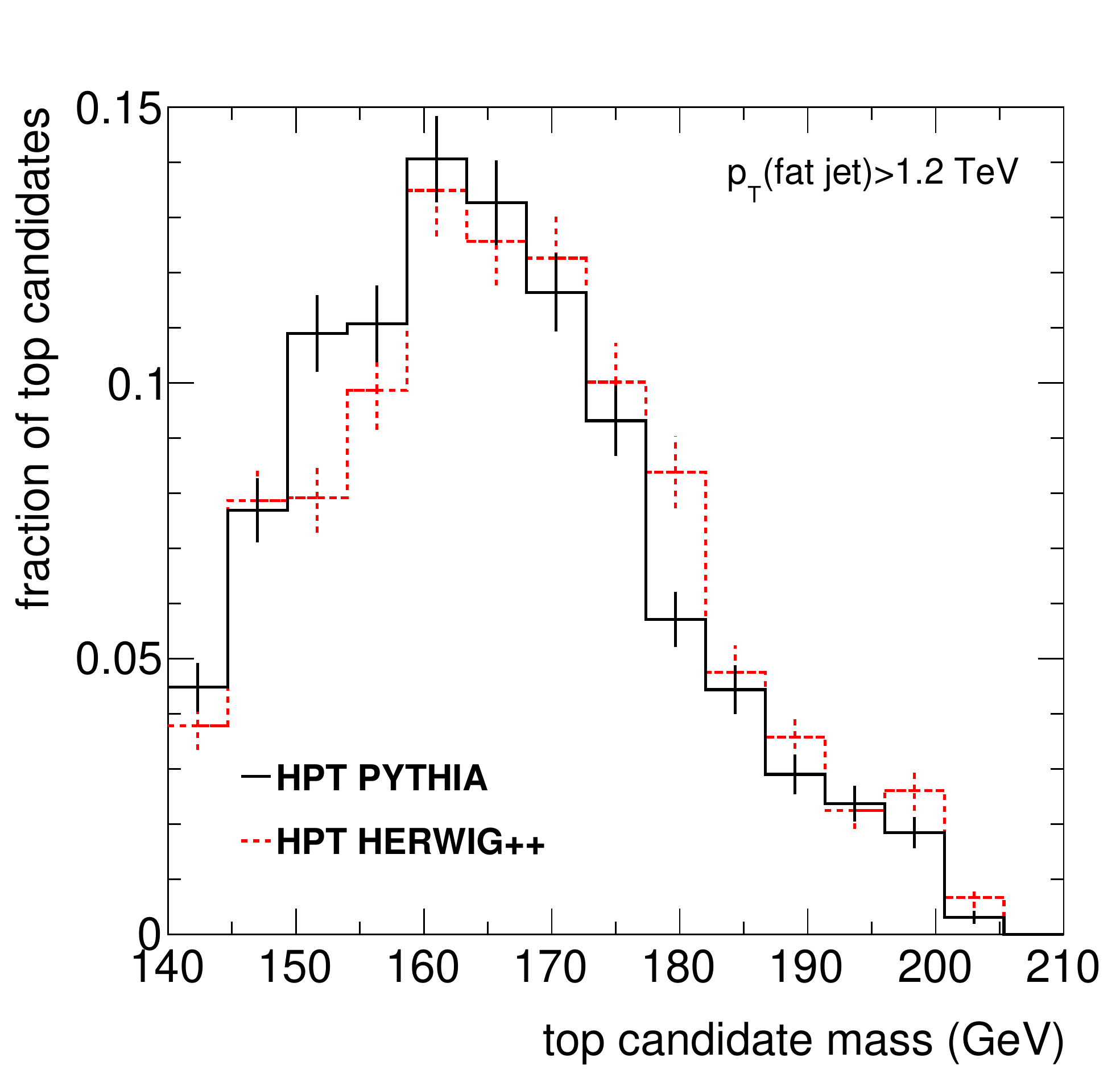}
\caption{Top quark mass reconstructed at the particle level with the \hptt using \ttbar events generated with
\pythia and \herwigpp for fat jets with $\pt>1.2$~TeV.}
\label{fig:herwig}
\end{figure}

The \hptt sensitivity to the imperfect knowledge of charged particle production, i.e., the hadronization model,
is small. Measurements of the jet fragmentation function and comparisons with different
generators have been reported in~\cite{Aad:2011sc}.
The difference between the string-model \cite{Andersson:1983ia} based \pythia 8 and the cluster-model \cite{Webber:1983if} based HERWIG++ 2.5~\cite{Bahr:2008pv} gives a conservative estimate
of the difference in charged particle production.
The efficiency for tagging fat particle jets with $\pt>1.2$~TeV is shown in \tabref{eff} for
two samples of \ttbar events with top quark $\pt>1$~TeV, one generated with \pythia~8
and one with \herwigpp~2.5. The efficiencies are compatible within the relative statistical uncertainty
of $3\%$. The top quark candidate mass, reconstructed at the particle level, is compared in \figref{herwig}.
The average mass from \herwigpp is larger by only 1.7~GeV.

\begin{table}[ht!]
  \begin{center}
    \begin{tabular}{|c|c|c|}
      \hline
      efficiency for tagging         & \pythia 8 & HERWIG++ 2.5 \\
      \hline
      the leading \pt fat jet     & $27.0(5)\%$ & $28.2(5)\%$ \\
      the sub-leading \pt fat jet & $18.1(4)\%$ & $18.6(5)\%$ \\
      both fat jets               & $4.9(1)\%$ & $5.3(2)\%$ \\
      \hline
    \end{tabular}
    \caption{The efficiency for tagging fat particle jets with $\pt>1.2$~TeV
             for two samples of \ttbar events with top quark $\pt>1$~TeV.}
  \label{tab:eff}
  \end{center}
\end{table}

\section{Reconstructing heavy resonances at the LHC}
\label{sec:res}

In the following we discuss top taggers in the context of detecting a
leptophobic topcolor \Zprime boson that decays to two top quarks~\cite{Harris:1999ya}.
The width of the resonance is set to $\Gamma_\Zprime/m_\Zprime= 3.2\%$.
We choose two mass points, $m_\Zprime=3~\mathrm{TeV}$ and $m_\Zprime=5~\mathrm{TeV}$,
for which the production cross sections in $pp$ collisions at $\sqrt{s}=14$~TeV
are $3.5$~fb and $97$~ab, respectively.

\begin{figure}[ht]
\includegraphics[width=3.0in]{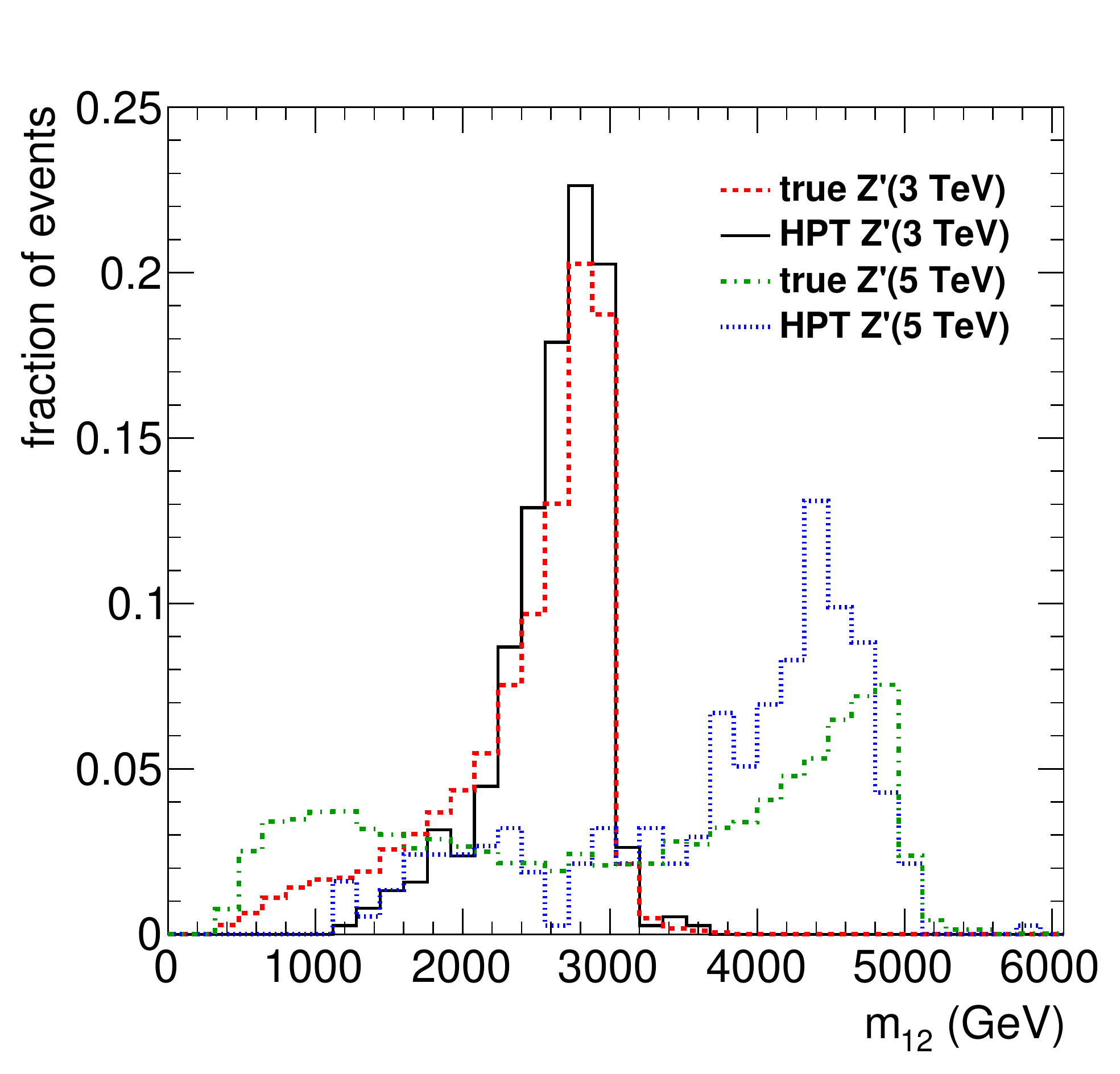}
\caption{Invariant mass of the two top quarks from the decay $\Zprime \rightarrow \ttbar$ (`true \Zprime') after
QCD radiation for two \Zprime masses and the corresponding reconstructed distributions when using the \hptt.}
\label{fig:trueZ}
\end{figure}

Top taggers imposing requirements on
the top quark mass and/or the \W boson mass reconstruct
on-shell top quarks right before they decay. Therefore, radiation off the
top quark, while necessary to reconstruct the $Z'$ resonance, is
discarded. Particularly for heavy \Zprime bosons, which result in highly
boosted top quarks, gluon radiation is not unlikely, as can be seen from
\figref{trueZ} which shows the invariant mass of the two top quarks after
QCD radiation.
Events with $m_{12} \leq 4$~TeV
amount to $2/3$ of the production cross section of a 5~TeV \Zprime boson.
Those events require a refined reconstruction strategy beyond the simple double top-tag
discussed here.

\begin{figure}[ht]
\subfigure[]{
   \includegraphics[width=3.0in]{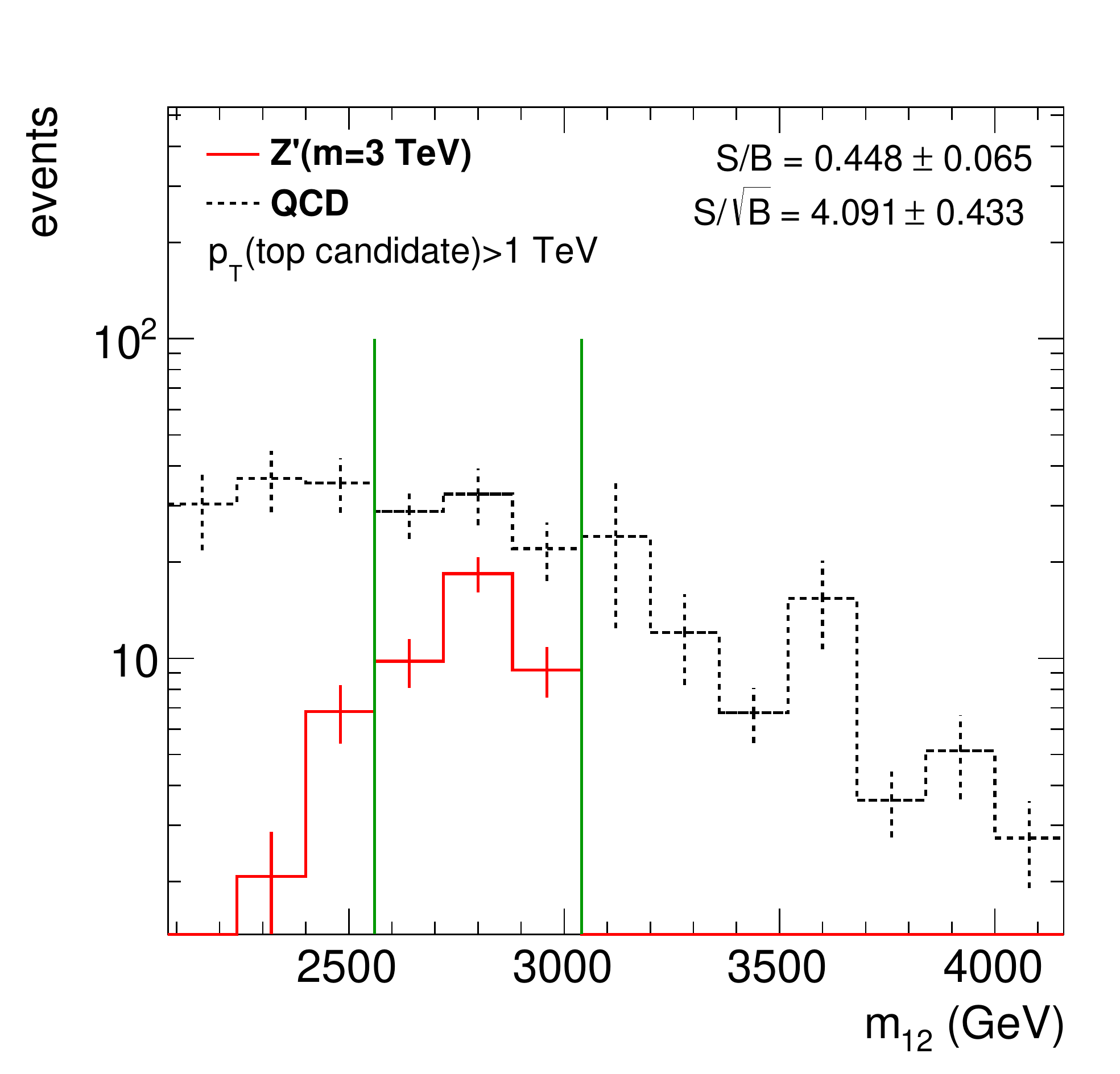}
}
\subfigure[]{
   \includegraphics[width=3.0in]{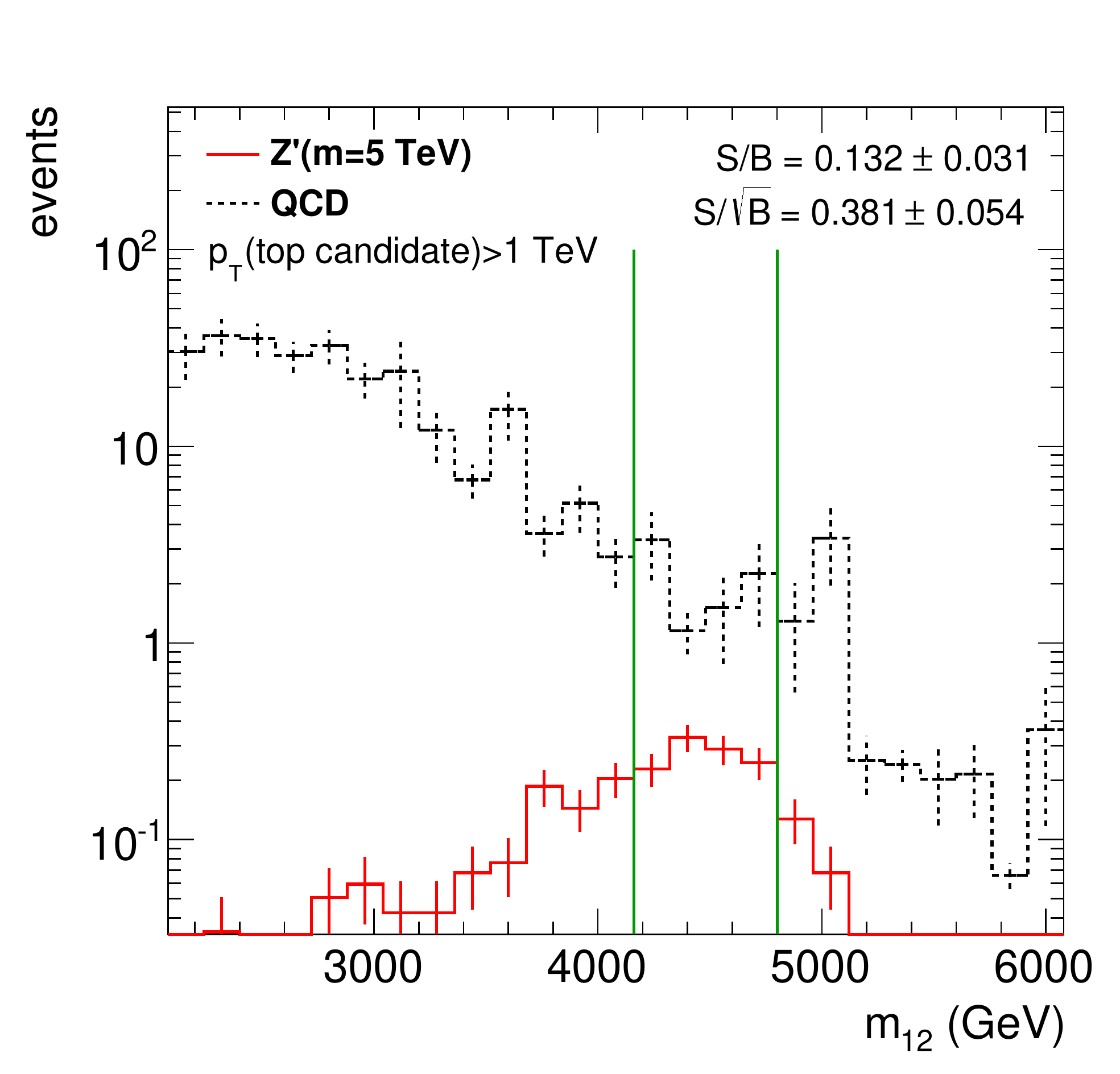}
}
\caption{Invariant di-top mass, reconstructed with the \hptt, from 300~fb$^{-1}$
of decays of \Zprime bosons of mass a) 3~TeV and b) 5~TeV, produced in $pp$ collisions
at $\sqrt{s}=14$~TeV.
Also shown is the background from QCD dijet production. The signal to noise ratio
$S/B$ and the significance $S/\sqrt{B}$ are given for the indicated mass window.}
\label{fig:recinvmass}
\end{figure}

The reconstructed di-top invariant mass is shown in \figref{recinvmass} for
the \Zprime signal and QCD dijet production, which constitutes the most
important background (\ttbar production is smaller by a factor of $\approx 0.1$)
for 300~fb$^{-1}$ of $pp$ collisions at a center-of-mass energy of 14~TeV.
The only imposed requirement is that there be at least two tagged \CamKt $R=0.8$ jets in the
event. The plotted quantity $m_{12}$ is the invariant mass of the two leading \pt
top quark candidates.
Based on the expected position of the signal, we have defined mass windows, in
which we compare the number of signal ($S$) and background events ($B$).
For the 3~TeV \Zprime boson, we find a signal-to-background ratio $S/B = 0.45(7)$
and a significance $S/\sqrt{B} = 4.1(4)$ in the window $2560<m_{12}<3040$~GeV.
The discovery of such a \Zprime boson with the \hptt is therefore within
reach. The uncertainties are statistical and dominated by the finite number of simulated background events.
For comparison, with the same generated events, the significance when using the \htt is only $3.3(3)$ and
the difference to the \hptt results directly from the different fat jet tagging
efficiencies shown in \figref{eff_fj}.
There is no sensitivity to a 5~TeV \Zprime boson,  with $S/B = 0.13(3)$ and
$S/\sqrt{B} = 0.38(5)$ in the window $4160<m_{12}<4800$~GeV,
because the background level is too high.
The sensitivity might be improved by applying $b$-quark tagging if the related
systematic uncertainties are small enough at high \pt.

\section{Summary and outlook}
\label{sec:outlook}
Traditional top quark finding algorithms, that are based on identifying the
3-prong hadronic decay structure, fail when the decay products can no longer be resolved.
For calorimeter granularities of $0.1\times 0.1$ we find this merging of particle jets
to start at top quark transverse momenta of $\approx 1.2$~TeV.

We propose the \hptt, a new algorithm to find boosted top quarks with transverse
momentum $\pt>1$~TeV, that combines track and calorimeter information. The finer
spatial resolution of tracking detectors allows the separation of close-by particle jets
that would merge in the calorimeter.
We have shown that with the \hptt, a \Zprime boson of mass 3~TeV is within
discovery reach when using 300~fb$^{-1}$ of 14~TeV LHC data.

Including photons, measured in a finely grained electromagnetic
calorimeter, in the \hptt algorithm could improve the performance.
For heavy resonances, the effect of QCD radiation off the top quarks becomes
important and a possible way to reconstruct the resonance mass is to include
jets in addition to the two top quark candidates.

While the HPTTopTagger has a smaller tagging efficiency for
top quarks with $p_T<1~\mathrm{TeV}$ compared to standard tagging approaches using subjets,
it performs better for highly boosted top quarks. We point out that it is
straightforward to combine the HPTTopTagger approach with any subjet-based
top tagger, particularly with the HEPTopTagger, to obtain an improved tagging
performance over a large $p_T$ range.

\begin{acknowledgments}

This work was funded in the UK by STFC. We thank the Perimeter Institute
and the University of Oregon for hospitality. We thank Simone Marzani for valuable discussions.

\end{acknowledgments}

\end{document}